\documentclass[aps,prc,twocolumn,showpacs,floatfix]{revtex4-1}

\usepackage{amssymb}
\usepackage{amsmath,bm}
\usepackage[normalem]{ulem}
\usepackage[dvips]{color}
\usepackage{booktabs}
\usepackage{graphicx}
\usepackage{float}
\usepackage{multirow}
\usepackage{lineno}

\usepackage[normalem]{ulem}

\usepackage[usenames, dvipsnames]{xcolor}
\usepackage{tensor}

\renewcommand{\sout}{\bgroup \color{red} \ULdepth=-.5ex \ULset}

\begin{document}

\title{Production of light antinuclei in $pp$ collisions by dynamical coalescence and their fluxes in cosmic rays near earth}
\author{Tianhao Shao$^{1}$}
\author{Jinhui Chen$^{1,2}$}
\author{Yu-Gang Ma$^{1,2}$}
\author{Zhangbu Xu$^{3}$}

\affiliation{$^1$Key Laboratory of Nuclear Physics and Ion-beam Application (MOE), Institute of Modern Physics, Fudan University, Shanghai 200433, China}
\affiliation{$^2$Shanghai Research Center for Theoretical Nuclear Physics, NSFC and Fudan University, Shanghai 200438, China}
\affiliation{$^3$Brookhaven National Laboratory, Upton, New York 11973, USA}
\date{\today}

\begin{abstract}
Light antinucleus yields are calculated in a multiphase transport model (AMPT) coupled with a dynamical coalescence model. The model is tuned to reproduce the transverse momentum and rapidity distributions of antiproton in $pp$ collisions at $\rm \sqrt{s}$ = 7.7 GeV to 7 TeV. By applying a widely used cosmic ray propagation model, the antinucleus fluxes near the earth are estimated over a broad range of kinetic energies. Our result on the antideuteron flux is consistent with the calculations in the literature, while our upper limit on the antihelium-3 flux sits in-between calculations in the field that span an order of magnitude in its value. This study suggests that further accurate estimation of secondary antihelium-3 flux could be improved with more ground-based experiments and model simulations. Most importantly, our value of antinucleus background from hadronic source is far below the projected sensitivity of AMS-02 with 5 years of integration time, which supports the idea of searching for new physics by measurements of light antinuclei in upcoming decade.  
\end{abstract}

\maketitle

\section{Introduction}
Searching for antinuclei in cosmic rays, which may be produced in the annihilation or decay of dark matter but are rarely produced in hadronic interactions, is among some of the promising breakthroughs in the dark matter searches~\cite{Donato:1999gy}. 
Experiments such as AMS-02~\cite{AMS:2021nhj}, DAMPE~\cite{DAMPE:2017cev} or GAPS~\cite{Aramaki:2014oda} with their detectors on the earth orbit or balloon flying in the sky, have been hunting for their signal in our galaxy. The flux of antiproton in cosmic rays has been measured~\cite{AMS:2016oqu,Adriani:2008zq}. However, no event of antideuteron has ever been detected. To date, AMS-02 has reported the observation of eight antihelium candidate events with rigidity (momentum per charge) below 50 GV, from a sample of 700 million selected helium events~\cite{AMSantihe}. Among reported antihelium events, six have masses in the range of antihelium-3, and two have masses in the range of antihelium-4. Although the detection of antihelium in cosmic rays has not been confirmed and officially published except in the earth-based heavy-ion experiments~\cite{STAR:2011eej,ALICE:2017jmf}, it has prompted many theoretical works on its implications for dark matter models and the predicted antideuteron and antiproton fluxes.

In order to assess whether any antinucleus candidates are from the dark matter source, their background from standard astrophysical and hadronic processes should be precisely evaluated~\cite{Duperray:2005si,vonDoetinchem:2020vbj}. In these processes, antinuclei are produced when cosmic-ray protons or antiprotons interact with the interstellar matter (ISM)~\cite{Duperray:2005si}. Since the majority matters in interstellar gas are hydrogens, these processes are just like antinuclei production in $pp$ collisions. It is crucial to distinguish secondary antinuclei from the signal candidates due to the dark matter. To obtain the cross sections for production of antinuclei in $pp$ collisions, the coalescence model together with the kinematic distributions from Monte Carlo generators are usually used. The flux of antinuclei near the earth from these processes can be simulated after their propagation in the galaxy. Several model calculations~\cite{Kachelriess:2020uoh,Blum:2017qnn,Poulin:2018wzu,Ding:2018wyi,Shukla:2020bql} have been done by following this procedure to predict the flux of antinuclei due to backgrounds from cosmic rays. The obtained flux can be compared to that from the antihelium candidate events in AMS-02 data. Reference~\cite{Blum:2017qnn} predicted that this background of antihelium-3 flux is 1 or 2 orders of magnitude higher than most earlier estimations, which may indicate that the antihelium candidate events in AMS-02 data can come from the collisions of background cosmic rays. However, these calculations usually use a naive momentum space coalescence model. In this model, the cross section of a nucleus with atomic mass number $A$ is calculated directly from the cross sections of its constituent nucleons:
\begin{align}\label{coal}
    E_A\frac{{\rm d}^3 \sigma_A}{{\rm d}k_A^{3}} = B_A \times \left(E_p\frac{{\rm d}^3 \sigma_p}{{\rm d} k_p^{3}} \right)^Z \times \left(E_n\frac{{\rm d}^3 \sigma_n}{{\rm d} k_n^{3}} \right)^{N},
\end{align}
where $Z$ and $N$ are, respectively, the proton number and the neutron number of nucleus $A$, and $k_A$, $k_p$, $k_n$ are the correspondingly momenta. Eq.~(\ref{coal}) relies on the coalescence factor $B_A$, which is usually taken from measurements in $pp$ collisions at LHC energies~\cite{ALICE:2017xrp}. Because of the steep energy spectrum of cosmic rays~\cite{AMS:2015tnn,DAMPE:2019gys}, a majority of the cosmic-ray protons are in lower energies. Thus, it is important to study the production of antinuclei at low energy $pp$ collisions. Recently, the NA61/SHINE collaboration has measured the antiprotons momentum spectra of antiparticles from $pp$ collisions at $\sqrt{\rm s}$ = 7.7 GeV to 17.3 GeV~\cite{NA61SHINE:2017fne}. It provides an important dataset to calibrate the momentum spectra of antinuclei from antinucleon coalescence in low energy $pp$ collisions.

In this work, we calculate the fluxes of antiproton, antideuteron, and antihelium-3 in cosmic rays. A multiphase transport model (AMPT)~\cite{Lin:2004en} is applied to generate the full phase-space density of antiproton and antineutron in $pp$ collisions. For the coalescence scenario, we use the dynamical coalescence model, which does not rely on the coalescence factor $B_A$, to calculate the energy distributions of antideuteron and antihelium-3. Their fluxes in cosmic ray are then determined by using the cosmic ray (CR) grammage model~\cite{Katz:2009yd} with the inclusion of the solar modulation.

\section{Antinucleus formation}
Most of the secondary antinuclei are produced in $pp$ collisions of the cosmic ray with the ISM. We use the AMPT model (v2.31t1) to generate the full phase-space density of antiprotons and antineutrons in such scenario. The AMPT model applies the kinetic theory approach to describe the evolution of heavy-ion collisions as it contains four main components: the fluctuating initial condition, partonic interactions, hadronization, and hadronic interactions~\cite{Lin:2004en,Lin:2021mdn}. It has been widely used to simulate the evolution of the dense matter created in high energy nuclear collisions. In particular, the string melting version of the AMPT model can describe well the anisotropic flows and particle correlations in collisions of $pp$, $pA$ or $AA$ systems at RHIC and LHC energies~\cite{Bzdak:2014dia,Zhang:2018ucx,Zhang:2020hww,Li:2021znq,Wang:2021xpv,Zhang:2021ygs,Zhao:2021bef,Wang:2022fwq,Zhu:2022dlc}. The key parameters in the Lund string fragmentation for generating the initial conditions for the AMPT model are set to $a_L$ = 0.5 and $b_L$ = 0.9 GeV$^{-2}$, respectively. To describe proton and antiproton production in low-multiplicity collisions, the parameter $r_{\rm BM}$, which controls the baryon-to-meson yield ratio, is set to 0.55 at LHC energies~\cite{He:2017tla,Shao:2020sqr}. This parameter is tuned to 0.75 in this study to describe the low energy data~\cite{NA61SHINE:2017fne}. 

\begin{figure}
    \centering
    \includegraphics[width=0.4\textwidth]{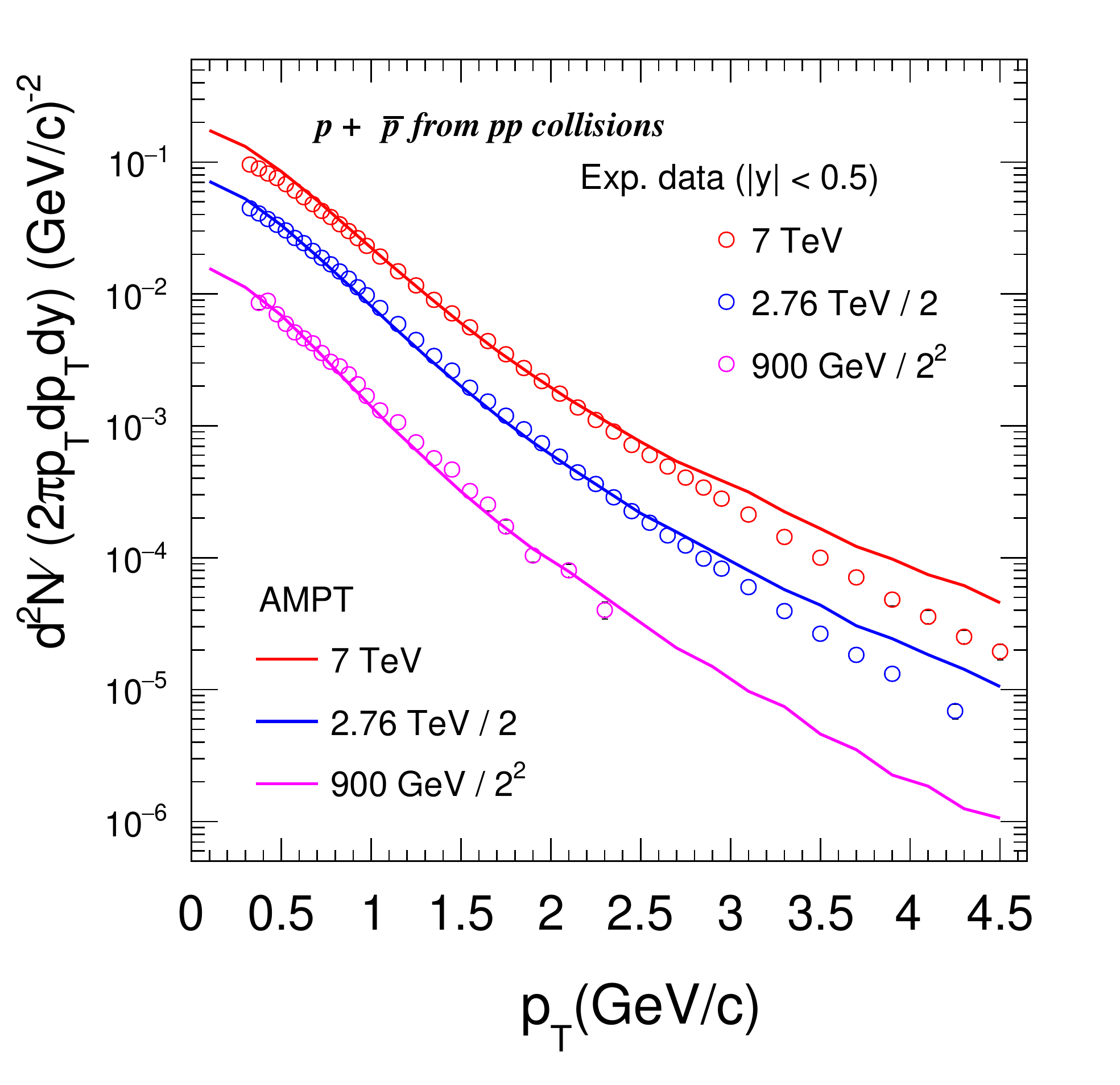}
    \caption{Transverse momentum distributions of proton and antiproton at midrapidity from $pp$ collisions at $\rm \sqrt{s}$ = 900 GeV, 2.76 TeV, and 7 TeV. Curves and points are AMPT model calculations and experimental data~\cite{ALICE:2011gmo,ALICE:2014juv,ALICE:2015ial}, respectively. For clarity, distributions for different energies are scaled by factor of two.}
    \label{fig:ppalice}
\end{figure}

Figure~\ref{fig:ppalice} shows the AMPT results for the transverse momentum $p_{\rm T}$ spectra of proton and antiproton at midrapidity ($\lvert y \rvert <$ 0.5) in $pp$ collisions at $\rm \sqrt{s}$ = 900 GeV, 2.76 TeV, and 7 TeV. Experimental data~\cite{ALICE:2011gmo,ALICE:2014juv,ALICE:2015ial} are shown for comparison. The AMPT model describes experimental data reasonably well up to $p_{\rm T}$$\approx$2.8 GeV/c. It overpredicts the data at higher $p_{\rm T}$ which has negligible contributions in our calculations to the total yields given the steep fallen spectra.

\begin{figure}
    \centering
    \includegraphics[width=0.4\textwidth]{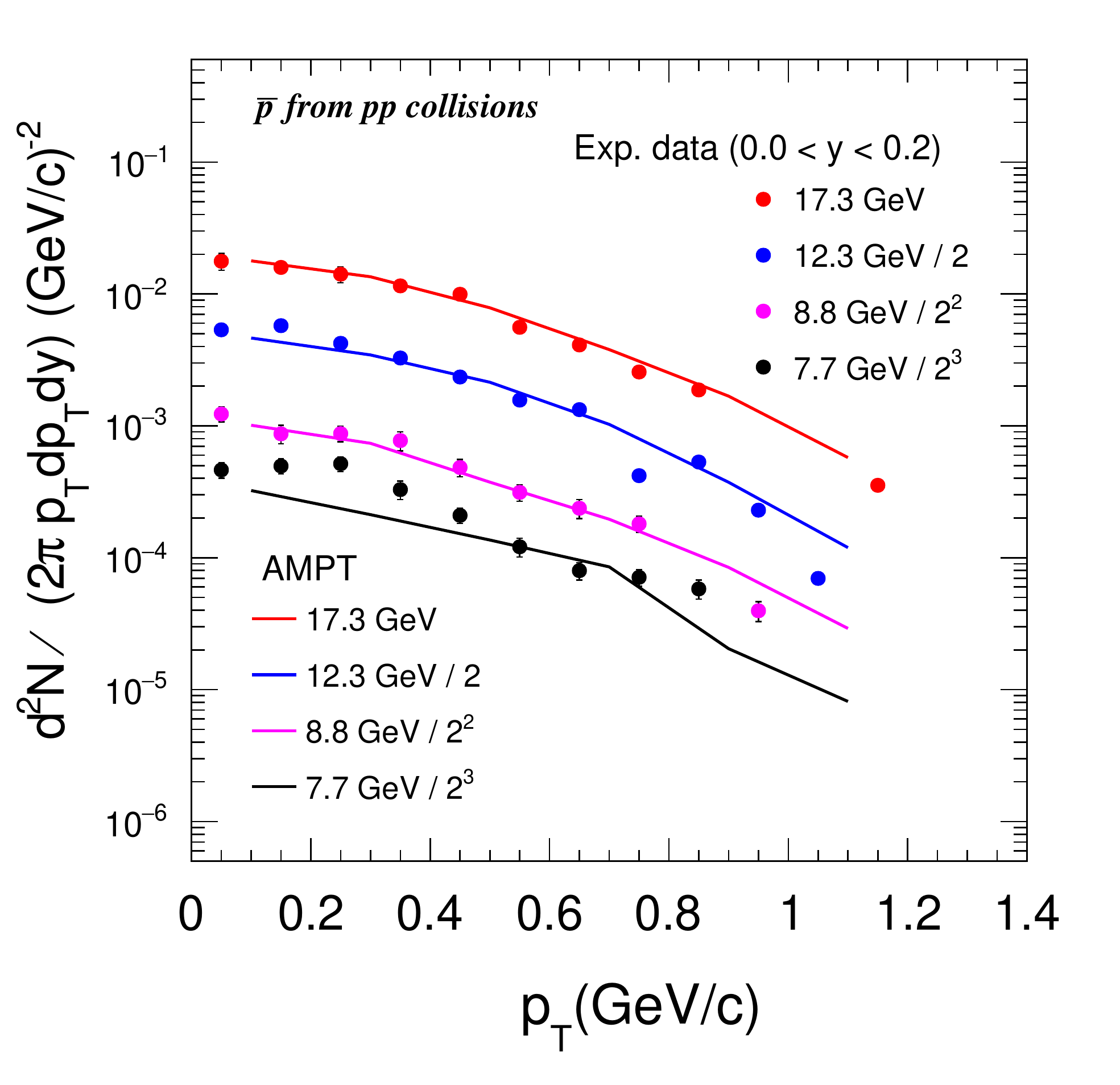}
    \caption{Similar to Fig.~\ref{fig:ppalice} but for antiproton $p_{\rm T}$ spectra in $pp$ collisions at $\rm \sqrt{s}$ = 7.7 GeV, 8.8 GeV, 12.3 GeV, and 17.3 GeV from AMPT model (solid lines) in comparison with experimental data (filled points)~\cite{NA61SHINE:2017fne}.}
    \label{fig:pp611}
\end{figure}

\begin{figure}
    \centering
    \includegraphics[width=0.4\textwidth]{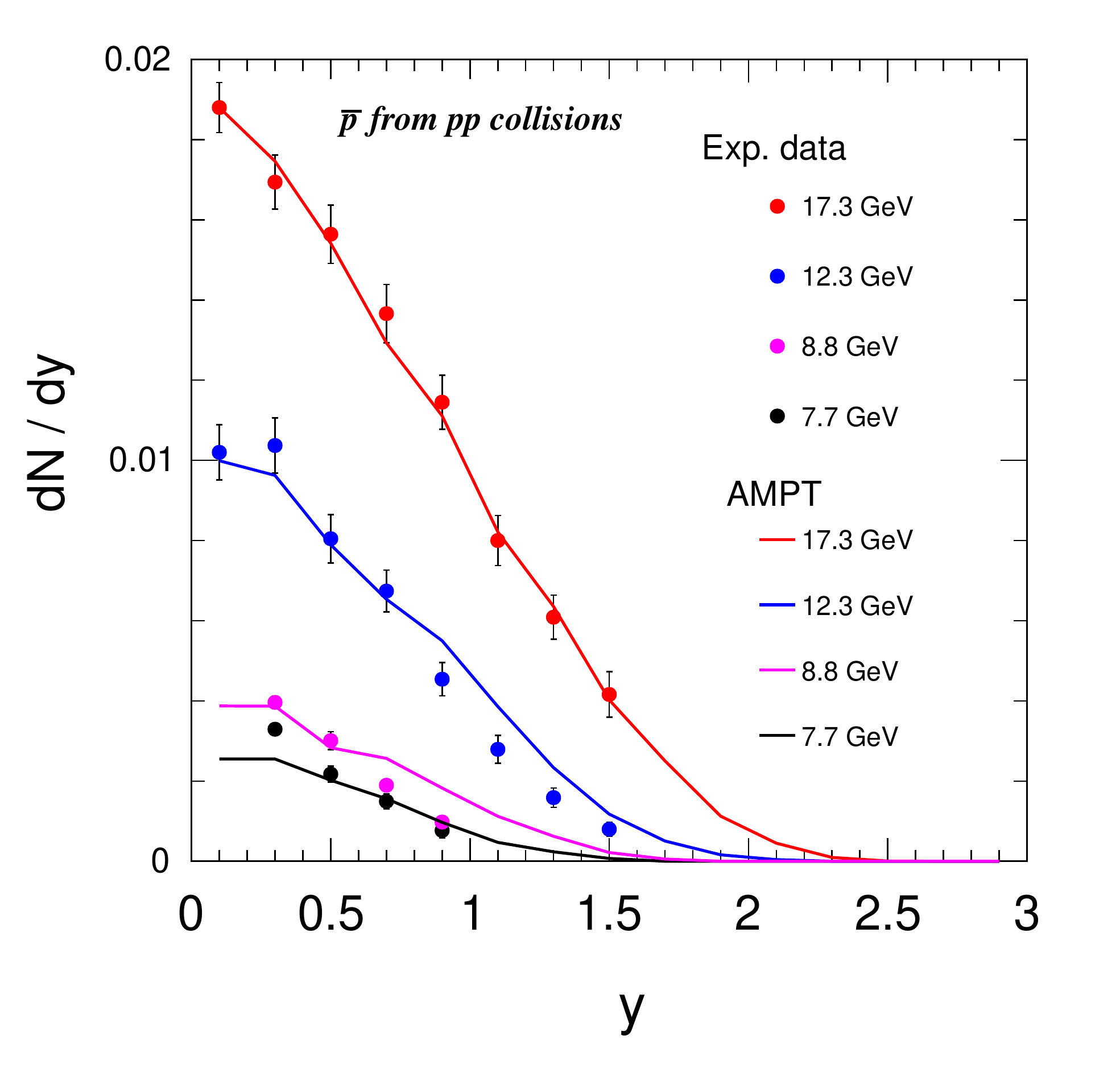}
    \caption{Rapidity density distributions of antiproton in $pp$ collisions at $\rm \sqrt{s}$ = 7.7 GeV, 8.8 GeV, 12.3 GeV, and 17.3 GeV from AMPT model (solid lines) in comparison with experimental data (filled points)~\cite{NA61SHINE:2017fne}.}
    \label{fig:pp612}
\end{figure}

Figures~\ref{fig:pp611} and~\ref{fig:pp612} show the AMPT results for the $p_{\rm T}$ and $y$ distributions of antiprotons in $pp$ collisions at $\rm \sqrt{s}$ = 7.7 GeV, 8.8 GeV, 12.3 GeV, and 17.3 GeV, together with the experimental results~\cite{NA61SHINE:2017fne}. In these lower collision energies, as aforementioned, the $r_{\rm BM}$ value in AMPT model is changed to 0.75 to match the charge particle rapidity density distributions. Thus the AMPT model can reproduce well the measured antiproton yield both in transverse and longitudinal direction. These momentum distributions together with the spatial distributions of antiprotons and antineutrons are used for calculating antinuclear production.

Productions of antideuteron and antihelium-3 are calculated by the dynamical coalescence model, which has been used successfully in describing the nuclei data in heavy-ion collisions~\cite{Zhang:2009ba,Xue:2012gx,Liu:2017rjm,Zhao:2020irc,Cheng:2021imy}. In this model the multiplicity of a nucleus with atomic mass number $A$ is calculated from
\begin{align}
N_{\rm A} =& g_{\rm A} \int {\rm d}\textbf{r}_{i_1}{\rm d}\textbf{q}_{i_1}\dots {\rm d}\textbf{r}_{i_{\rm A-1}}{\rm d}\textbf{q}_{i_{\rm A-1}} \nonumber\\ 
     &\times \langle\sum \rho_i^W({\rm d}\textbf{r}_{i_1}, {\rm d}\textbf{q}_{i_1}, \dots, {\rm d}\textbf{r}_{i_{\rm A-1}}, {\rm d}\textbf{q}_{i_{\rm A-1}}),
\end{align}
where $\textbf{r}$ and $\textbf{q}$ are the relative coordinates and momenta among nucleons in the light nucleus rest frame, $\rho_i^W$ is the Wigner phase-space density of the formed nucleus. $g_{\rm A} = (2J_{\rm A}+1)/2^{\rm A}$ is the statistical degeneracy factor for the nucleus, which is $3/4$ and $1/4$ for deuteron and helium-3, respectively.

The wave function of a deuteron can be taken to be:
\begin{align}
    \psi(\textbf{r}_1, \textbf{r}_2) = 1/(\pi\sigma_{\rm d}^2)^{3/4} \exp[-\textbf{r}^2/(2\sigma_{\rm d}^2)],
\end{align}
where $\textbf{r} = \textbf{r}_1 - \textbf{r}_2$, and $\sigma_{\rm d} = \frac{2}{\sqrt{3}}r_{\rm d}$ with $r_{\rm d} = 1.96~fm$ being its radius.
Then the Wigner phase-space density function of deuteron can be obtained as:
\begin{align}
    \rho_{\rm d}^W(\textbf{r}, \textbf{k}) =& \int\psi(\textbf{r}+\frac{\textbf{R}}{2})\psi^*(\textbf{r}+\frac{\textbf{R}}{2})\times\exp(-i\textbf{k}\cdot\textbf{R}){\rm d}^3\textbf{R} \nonumber\\
    =&~8\exp\left(-\frac{\rho^2}{\sigma_{\rm d}^2} - \sigma_{\rm d}^2\textbf{k}^2\right),
\end{align}
where $\textbf{k} = (\textbf{k}_1 - \textbf{k}_2)/\sqrt{2}$ and $\rho = (\textbf{r}_1 - \textbf{r}_2)/\sqrt{2}$.

For helium-3, the wave function is:
\begin{align}
    \psi(\textbf{r}_1, \textbf{r}_2, \textbf{r}_3) = (3\pi^2b^4)^{-3/4}\exp\left(-\frac{\rho^2+\lambda^2}{2b^2}\right),
\end{align}
The Wigner phase-space density function of helium-3 is:
\begin{align}
    \rho_{3}^W =& \int\psi(\rho+\frac{\textbf{R}_1}{2}, \lambda+\frac{\textbf{R}_2}{2})\psi^*(\rho-\frac{\textbf{R}_1}{2}, \lambda-\frac{\textbf{R}_2}{2}) \nonumber\\
    &~\times \exp(-i\textbf{k}_{\rho}\cdot\textbf{R}_1)\exp(-i\textbf{k}_{\lambda}\cdot\textbf{R}_2)3^{3/2}{\rm d}^3\textbf{R}_1{\rm d}^3\textbf{R}_2 \nonumber\\
    =&~8^2\exp\left(-\frac{\rho^2+\lambda^2}{b^2} - \textbf{k}_{\rho}^2b^2 - \textbf{k}_{\lambda}^2b^2\right),
\end{align}
where $\textbf{k}_{\rho}=\frac{1}{\sqrt{6}}(\textbf{r}_1 + \textbf{r}_2 - 2\textbf{r}_3)$, $\textbf{k}_{\lambda}=\frac{1}{\sqrt{6}}(\textbf{k}_1 + \textbf{k}_2 - 2\textbf{k}_3)$ and $b = r_{\rm ^3He} = 1.74~fm$, with $r_{\rm ^3He}$ being the radius of helium-3. 

\begin{figure}
    \centering
    \includegraphics[width=0.4\textwidth]{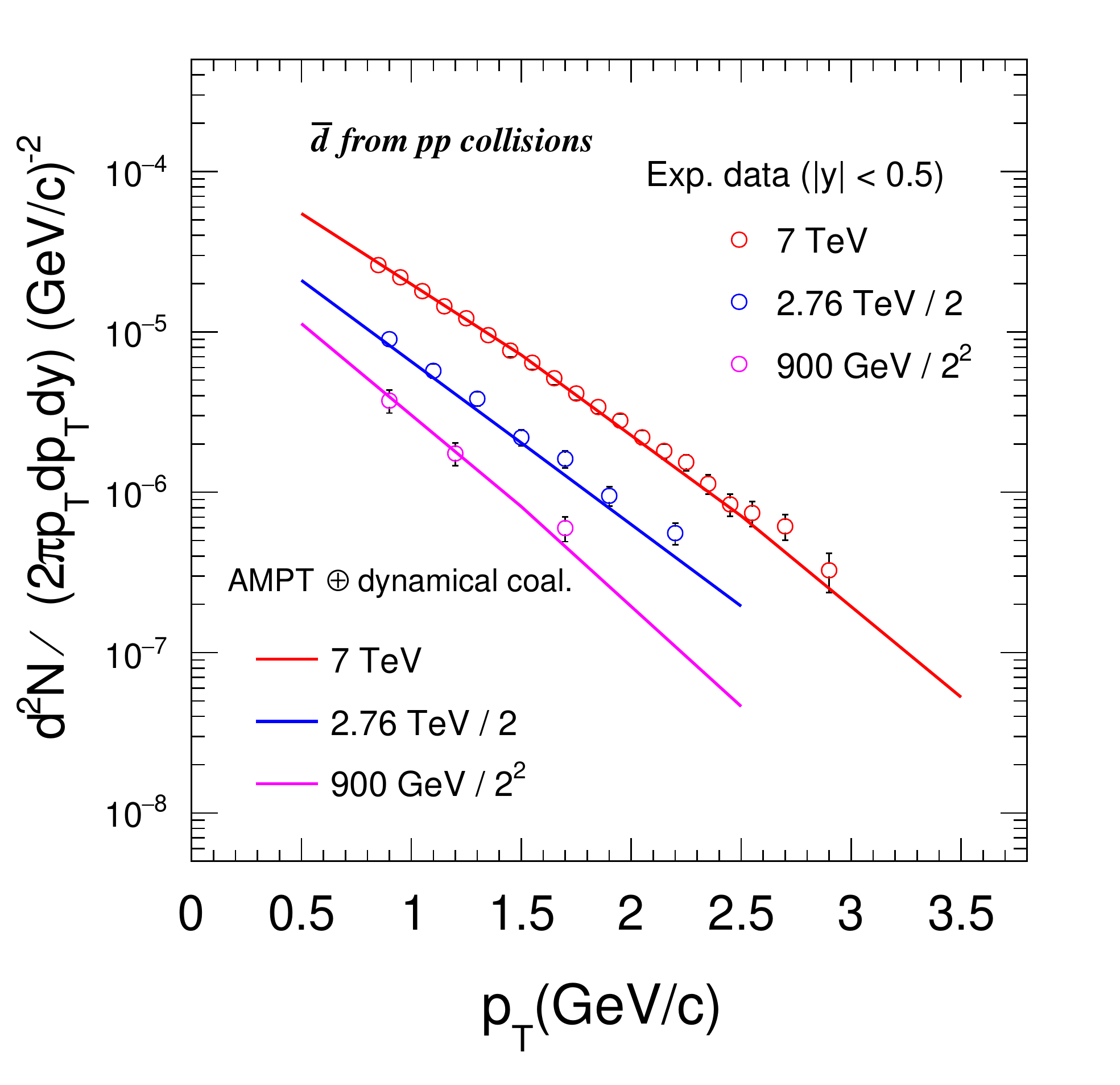}
    \caption{Similar to Fig.~\ref{fig:ppalice} but for antideuteron $p_{\rm T}$ spectra. Curves are results from AMPT coupled with a dynamical coalescence model, and points are experimental data~\cite{ALICE:2017xrp}.}
    \label{fig:dalice}
\end{figure}

\begin{figure}
    \centering
    \includegraphics[width=0.4\textwidth]{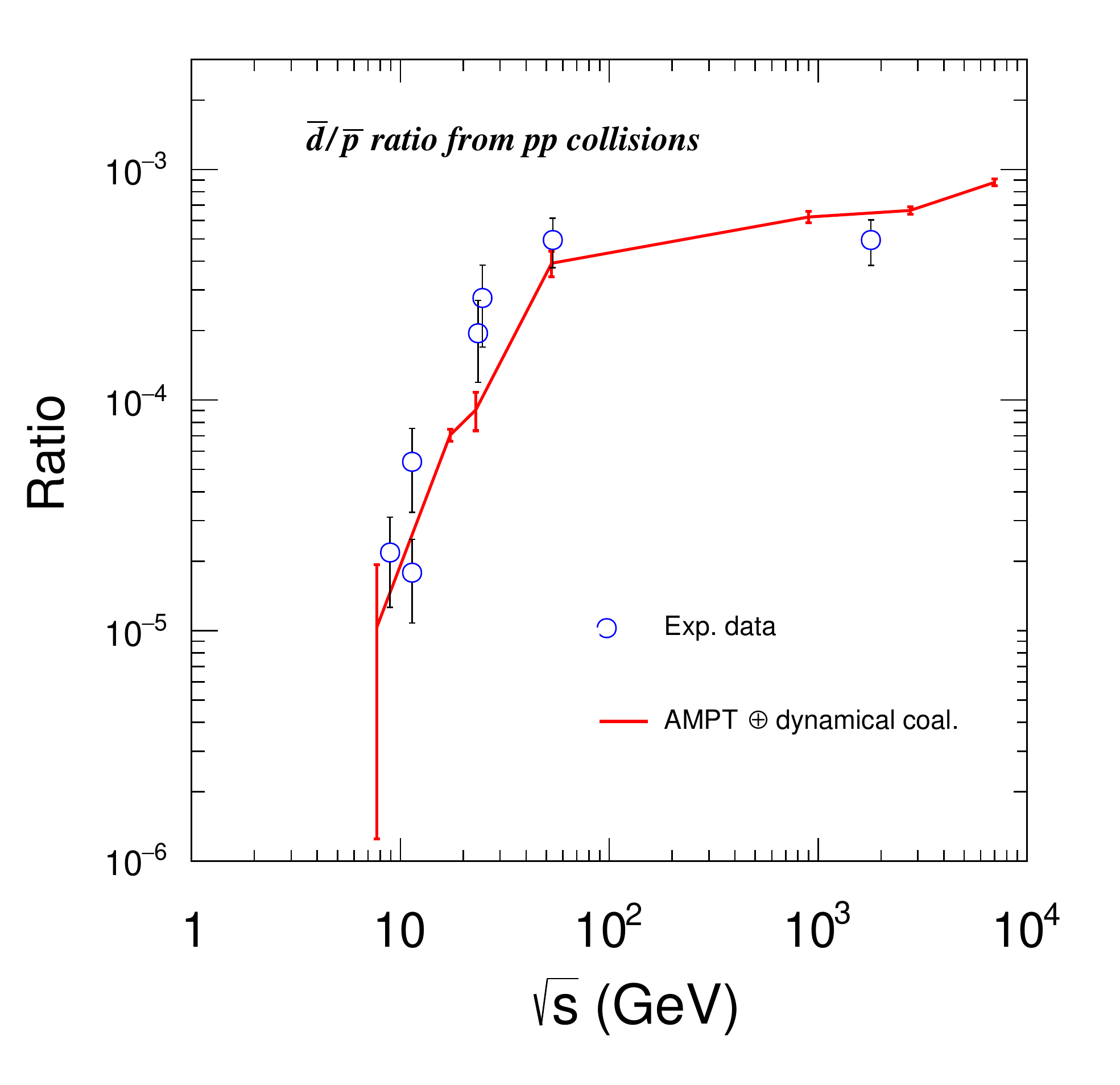}
    \caption{Collision energy $\rm \sqrt{s}$ dependence of the antideuteron yield to antiproton yield ratio. Curve represents the results from AMPT coupled with a dynamical coalescence model, and points are experimental data~\cite{IHEP-CERN:1969vgt,Appel:1974fs,Alper:1973my,Alexopoulos:2000jk}.}
    \label{fig:dp}
\end{figure}

These formulae and parameters are used in the calculations of antideuteron and antihelium-3. In our approach, antideuteron is formed by the coalescence of an antiproton and an antineutron with their phase-space information obtained from the AMPT model. Figure~\ref{fig:dalice} represents the $p_{\rm T}$ spectra of antideuteron by the dynamical coalescence model in $pp$ collisions at LHC energies. The AMPT model coupled with a dynamical coalescence model reproduces the antideuteron $p_{\rm T}$ spectra well~\cite{ALICE:2017xrp}. Figure~\ref{fig:dp} shows the yield ratio of antideuteron to antiproton over a broad range of collision energy $\rm \sqrt{s}$. It shows that the measured yield of antideuteron can also be reproduced at low energies. 

\begin{figure}
    \centering
    \includegraphics[width=0.4\textwidth]{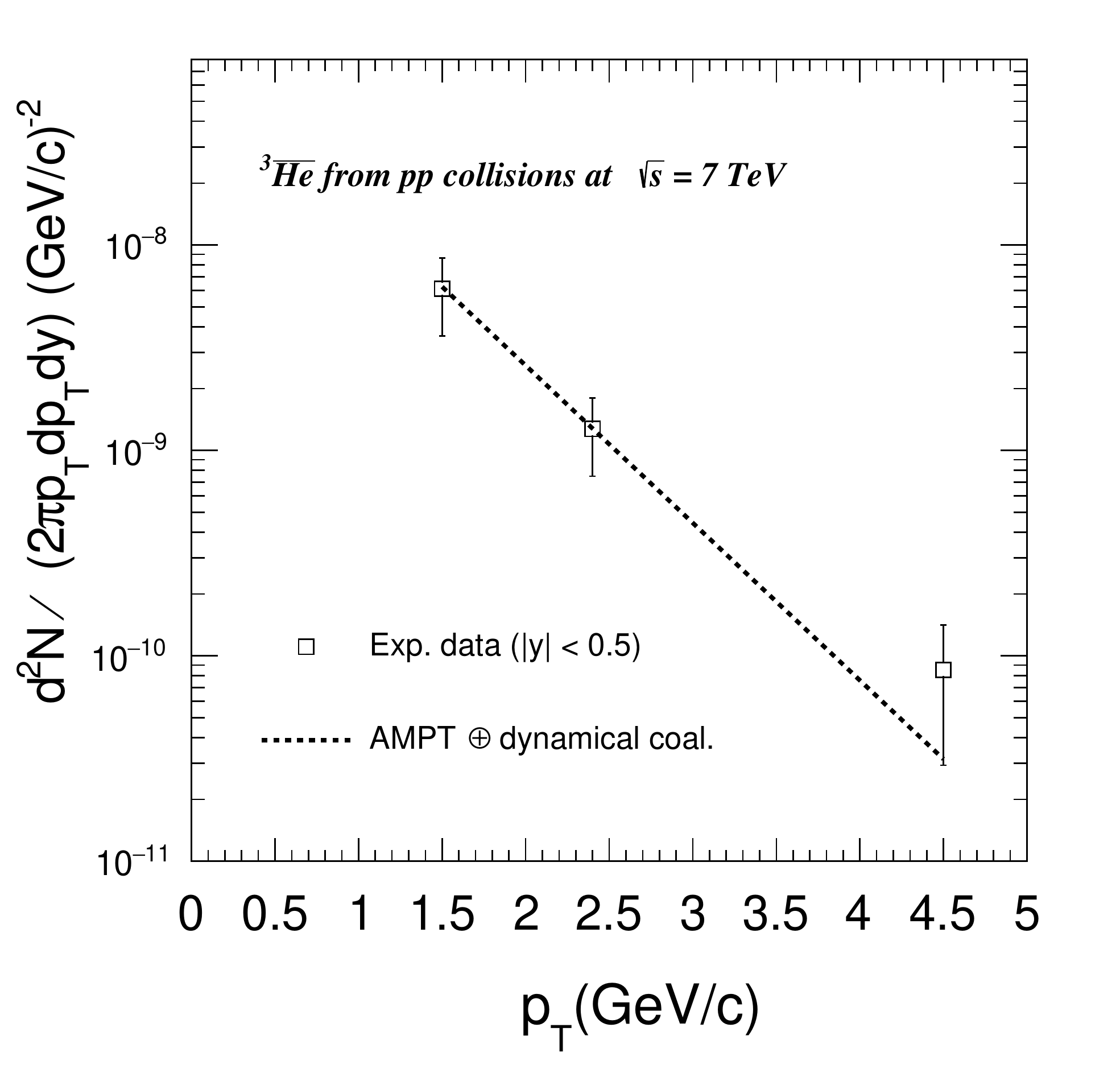}
    \caption{Transverse momentum spectrum of antihelium-3 at midrapidity from $pp$ collisions at $\rm \sqrt{s}$ = 7 TeV. Dotted curve represents the model calculation and points are experimental data~\cite{ALICE:2017xrp}.}
    \label{fig:he3alice}
\end{figure}

Figure~\ref{fig:he3alice} shows the $p_{\rm T}$ spectrum of antihelium-3 in $pp$ collisions at $\sqrt{\rm s}$ = 7 TeV from our calculation in comparison with experimental data~\cite{ALICE:2017xrp}. Our results reproduce the antihelium-3 $p_{\rm T}$ spectra at $\sqrt{\rm s}$ = 7 TeV. The $A$ = 3 antinucleus measurements in laboratory are limited. Measurements are usually performed in $pA$ or $AA$ reactions~\cite{Antipov:1971iq,STAR:2001pbk,STAR:2010gyg,ALICE:2015wav,Chen:2018tnh}. Data in $pp$ collisions are scarce and only available at TeV energies~\cite{ALICE:2017xrp,ALICE:2021ovi,ALICE:2021mfm}. We thus do not have the chance to test our model calculation in lower collision energies below TeV. We will discuss its impact in Section IV.

\section{Antinucleus propagation}
In this work we calculate the propagation of antinuclei in cosmic rays by the CR grammage model~\cite{Katz:2009yd,Blum:2017qnn}. It is assumed in this model that the local density of cosmic rays at a given rigidity (momentum per charge) is proportional to their local generation rate. In this CR grammage model without spallation loss, any local densities of two stable cosmic rays, 1 and 2, at a given rigidity are related by the equation:
\begin{align}
    \frac{n_{\rm 1}}{n_{\rm 2}} = \frac{Q_{\rm 1}}{Q_{\rm 2}} ,
    \label{eq1}
\end{align}
where $n_i$ and $Q_i$ are the density and the local production rate of the secondary $i$. $Q_i$ is given by
\begin{align}
    Q_i = \sum_{j\neq i}n_j \frac{\sigma_{j\to i}}{m_p}c\rho_{\rm ISM},
\end{align}
where $\sigma_{j\to i}$ is the production cross section of the secondary nucleus $i$ from the parent nucleus $j$, $c$ is the velocity of light and $\rho_{\rm ISM}$ is the density of ISM. The production rate from the source of antinuclei, which are from the collisions of primary protons in the cosmic ray with the ISM, can be written as 
\begin{align}
    Q_i = 4\pi n_{\rm ISM}\int_{T_{\rm min}}^{\infty}{\rm d}T_j\frac{{\rm d}\sigma_{j\to i}(T_j, T_i)}{{\rm d}T_i}\phi_j(T_j) , 
    \label{eq:source}
\end{align}
where $n_{\rm ISM}$ is the density of ISM gas. Here we use $n_{\rm H}$ = 1~cm$^{-3}$ for the hydrogen. The different cross section $\sigma_{j\to i}$ for antinuclei produced at different energy $T_i$ follows
\begin{align}
    \frac{d\sigma_{j\to i}(T_j, T_i)}{dT_i} = \sigma_{ji,\rm inel}\frac{dN_i(T_j,T_i)}{dT_i},
\end{align}
where $\sigma_{ji,\rm inel}$ is the total inelastic cross section for a proton with kinetic energy $T_j$ reacts with a `fixed-target' proton in ISM, and it can be obtained from the measurements in p+p collisions in the laboratory~\cite{ParticleDataGroup:2020ssz}. ${\rm d}N_i(T_j,T_i)/{\rm d}T_i$ is the kinetic energy distribution of secondary antinuclei which are calculated by the AMPT model coupled with a dynamical coalescence model as aforementioned. The $\phi_j(T_j)$ is the flux of cosmic ray protons which can be obtained by parametrizing the primary cosmic ray flux from space-based experimental measurements~\cite{AMS:2015tnn,DAMPE:2019gys}:
\begin{align}
    \phi(T) = aT^{-\gamma}\left(\frac{T}{T+b}\right)^c\prod_{i=1}^{N}f(T_{bi},\Delta y_i,s),
\end{align}
where
\begin{align}
    f(T_{b},\Delta y,s)=\left[1+\left(\frac{T}{T_b}\right)^s\right]^{\Delta \gamma/s}.
\end{align}
Following Ref.~\cite{Kachelriess:2020uoh}, we take parameters for fits of the proton flux with $N$ = 2, $a=26714~\rm m^2s~sr/(GeV/n)$, $b=0.49~\rm GeV/n$, $c=6.81$, $\gamma=2.88$, $T_{b1} = 343~\rm GeV/n$, $T_{b2} = 19503~\rm GeV/n$, $\Delta \gamma_1 = 0.265$, $\Delta \gamma_2 = -0.264$ and $s=5$.

Since particles propagate in the same manner, values of $Q_1/Q_2$ are the same for certain cosmic rays 1 and 2. Eq.~(\ref{eq1}) can then be rewritten as
\begin{align}
    n_i(\varepsilon)=\frac{Q_i(\varepsilon)}{c\rho_{\rm ISM}}X_{\rm esc}(\varepsilon/Z),
    \label{eq2}
\end{align}
where $\varepsilon$ is the energy of the particle and $Z$ is the proton number of a nucleus. $X_{\rm esc}$ is the grammage function that parameterizes the column density of target material traversed by the cosmic rays and is the same for all species. 

Taking into consideration of the spallation of the cosmic nuclei, we introduce the net production rates $\tilde{Q}_i$,
\begin{align}
    \tilde{Q}_i = \frac{Q_i}{c\rho_{\rm ISM}} - \frac{n_i\sigma_i}{m_p},
\end{align}
where $\sigma_i$ is the cross section for destruction of the cosmic ray per ISM nucleon and it is estimated approximately by $\sigma_i \approx 40A_i^{0.7}$ mb~\cite{Katz:2009yd}, with $A_i$ the mass number of the nucleus $i$. 
After the source term is replaced and the density of secondary $i$ is rewritten in Eq.~(\ref{eq2}) as 
\begin{align}
    n_i(\varepsilon)=\tilde{Q}_i X_{\rm esc}(\varepsilon/Z),
\end{align}
the density is then given by
\begin{align}
    n_i = \frac{X_{\rm esc}Q_i/(\rho_{\rm ISM}c)}{1+(\sigma_i/m_p)X_{\rm esc}}.
\end{align}
The parametrization of $X_{\rm esc}$ can be obtained by fitting to various experimental measurements of cosmic rays~\cite{Katz:2009yd},
\begin{align}
    X_{\rm esc} \approx 8.7\left(\frac{\varepsilon}{10Z~GeV}\right)^{-0.5} g\cdot {\rm cm}^{-2} .
\end{align}

The flux of the secondary antinuclei produced by cosmic ray proton interacting with ISM can be calculated with
\begin{align}
    \phi_i = v\cdot n_i / 4\pi = \beta c \cdot \frac{Q_i}{4\pi} \cdot \frac{X_{\rm esc}/(c\rho_{\rm ISM})}{1+(\sigma_i/m_p)X_{\rm esc}}.
\end{align}

The flux of a given cosmic ray can be modulated by the solar effects. Here we follow the classic formula~\cite{Webber:2003cj}:
\begin{align}
    J(E) = \frac{E^2-m^2}{(E+|Z|\Phi)^2-m^2}\cdot \phi_{i}(E+|Z\Phi|),
\end{align}
where $E$ and $m$ are the energy and mass of the nucleus, respectively, and $\phi_{i}$ is the flux without the solar modulation. The modulation parameter is set to $\Phi = 450$~MV~\cite{Webber:2003cj}.

\section{Results and discussions}

\begin{figure}
    \centering
    \includegraphics[width=0.4\textwidth]{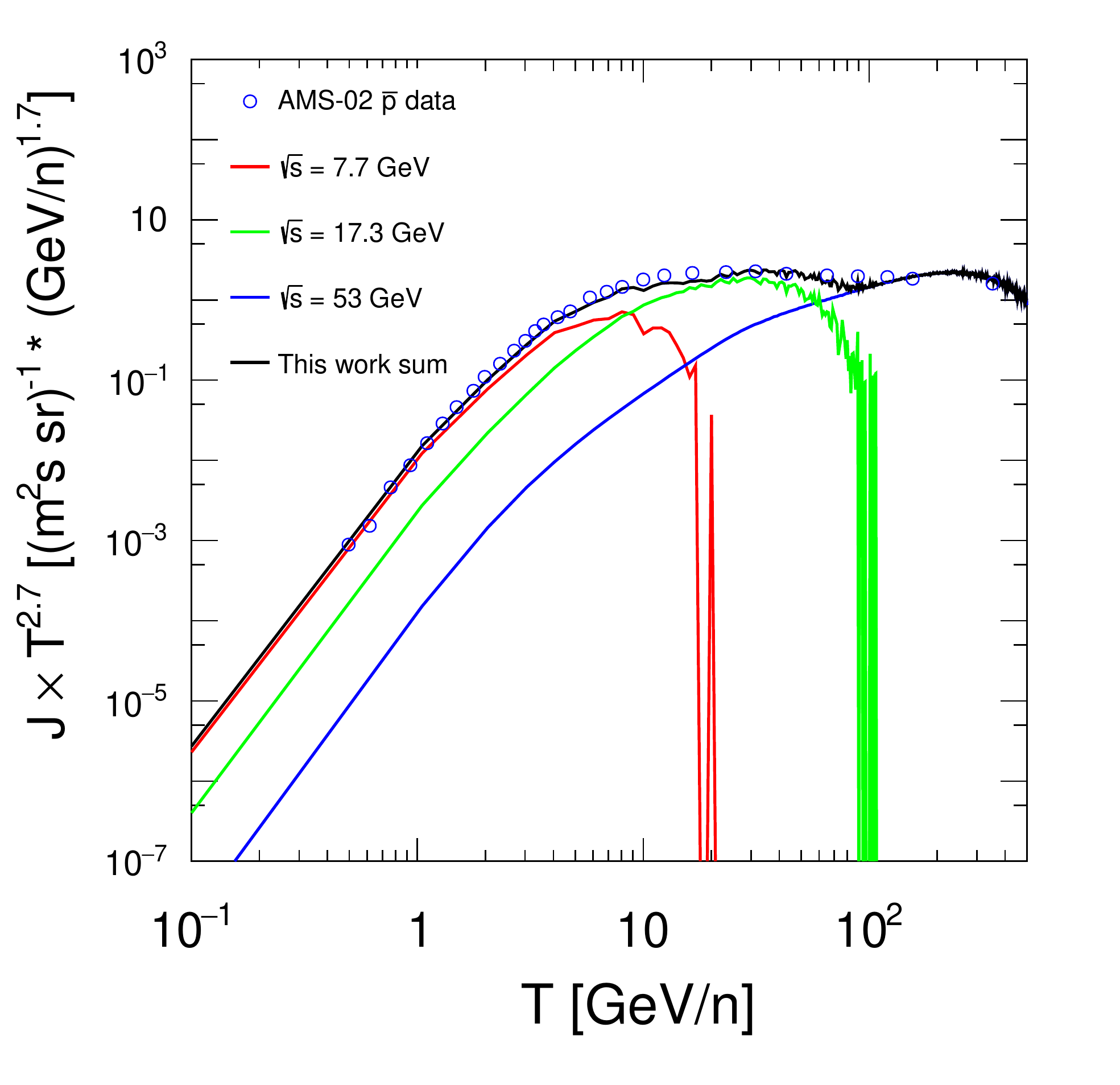}
    \caption{Flux of kinetic energy distribution of antiprotons multiplied by $T^{2.7}$ near earth produced from our calculation (lines) in comparison with space measurement (points)~\cite{AMS:2015tnn}.}
    \label{fig:pflux}
\end{figure}

According to Eq.~(\ref{eq:source}), the antinucleus production is from collisions between ISM and cosmic-ray protons with a continuous energy spectra. In our calculation, we divide the energy spectra into three bins with different center of mass energies of 7.7 GeV, 17.3 GeV, and 53 GeV, to represent the low, moderate and high energy regions, respectively. Figure~\ref{fig:pflux} shows our calculations of the antiproton flux multiplied by $T^{2.7}$ near the earth from $pp$ collisions in cosmic rays. It is seen that the $\rm \sqrt{s}$ = 53 GeV collision describes the high energy antiproton flux, while for the low kinetic energy region, $\rm \sqrt{s}$ = 7.7 GeV generates 2 order of magnitude higher flux and describes the data well. The sum of different collision energies reproduces the antiproton flux measured in space~\cite{AMS:2015tnn}. This serves as an important validation to support the predictive power of our current calculation method. It also suggests that indirect dark matter searches using antiprotons suffer from relatively high astrophysical background. Searching for a dark matter signal in the antiproton flux is thus challenging.

\begin{figure}
    \centering
    \includegraphics[width=0.4\textwidth]{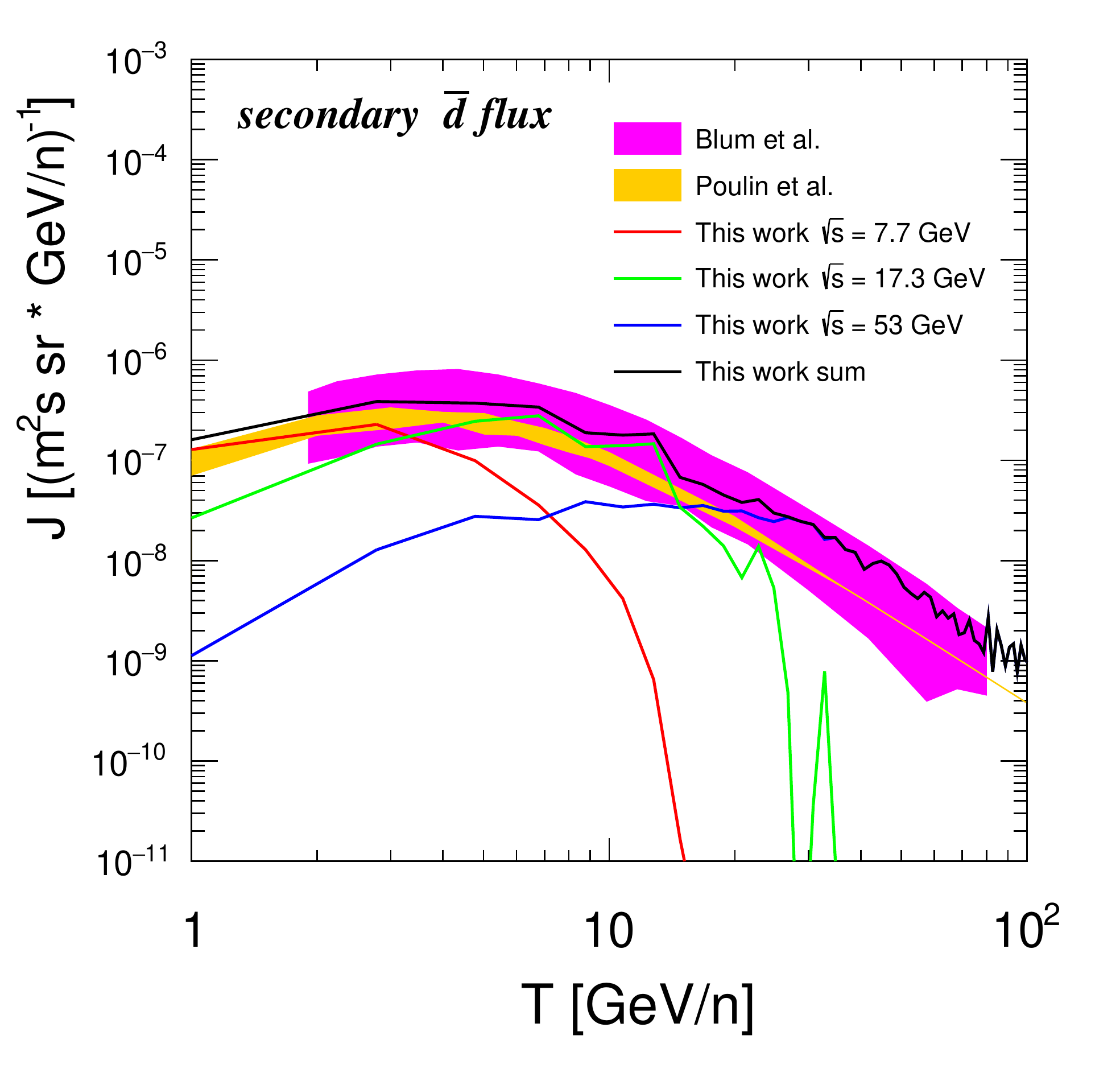}
    \caption{Flux of kinetic energy distribution of antideuteron near earth produced from our calculation (lines) in comparison with other calculations (band). The red, green, and blue lines show our results in $pp$ collisions at $\rm \sqrt{s}$ = 7.7 GeV, 17.3 GeV, and 53 GeV, respectively, and the black line is a sum of the three calculations. Pink band and orange band present results from Ref.~\cite{Blum:2017qnn}, and Ref.~\cite{Poulin:2018wzu}, respectively.}
    \label{fig:dflux}
\end{figure}

Figure~\ref{fig:dflux} shows the antideuteron flux from the AMPT model coupled with a dynamical coalescence model. The flux at each collision energy shows a similar kinetic energy distribution as the flux of antiprotons [c.f. Fig.~\ref{fig:pflux}]. Our results are consistent with estimates in Refs.~\cite{Blum:2017qnn} and~\cite{Poulin:2018wzu} within uncertainties. The maximum flux of antideuteron from $pp$ collisions in cosmic rays corresponds to a kinetic energy of about 6 GeV/n. Since the kinetic energy distribution of predicted dark matter signal also drops fast~\cite{vonDoetinchem:2020vbj}, our calculation suggests that it is optimal to hunt for antideuteron in a lower kinetic energy region. 

\begin{figure}
    \centering
    \includegraphics[width=0.4\textwidth]{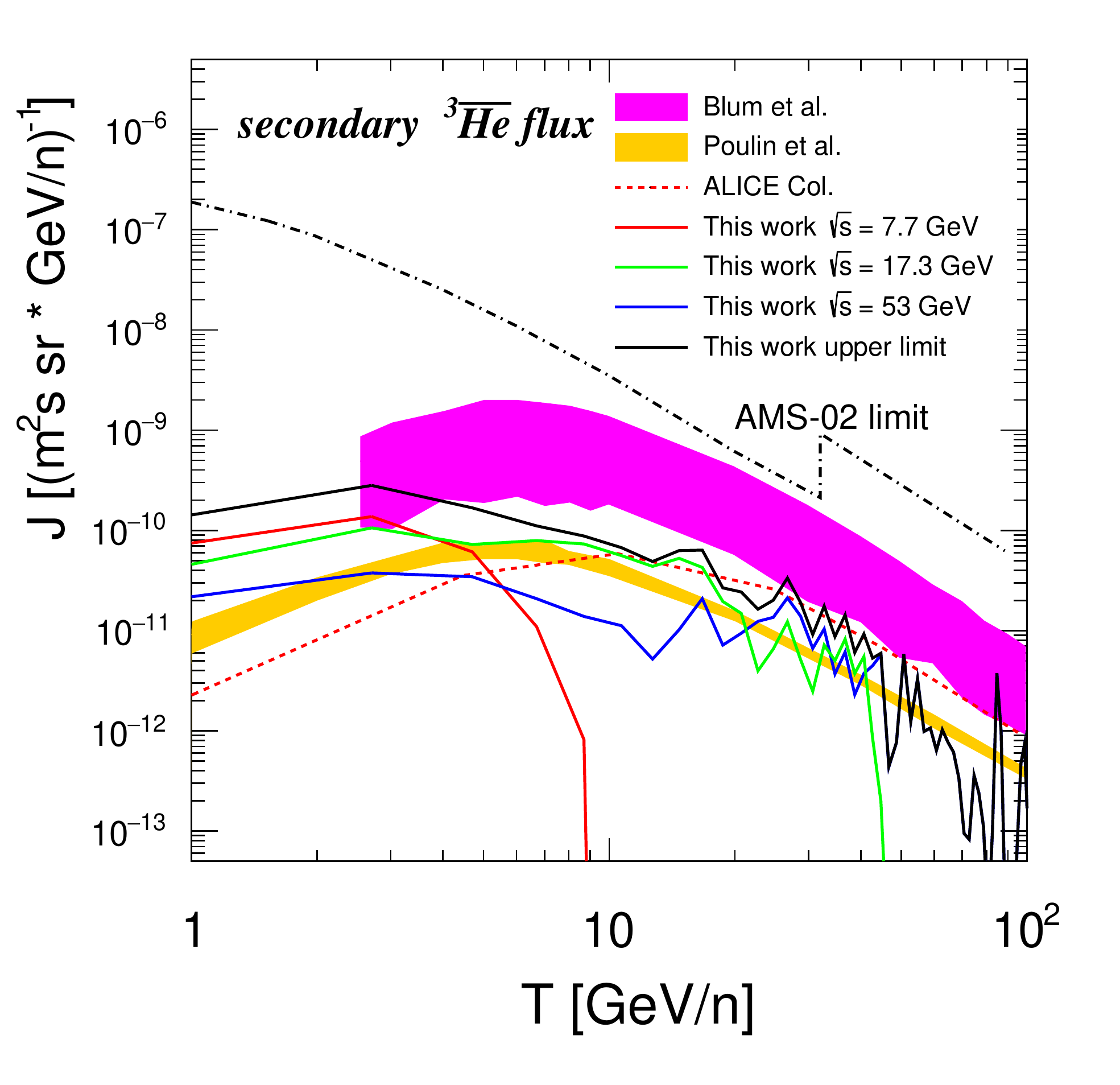}
    \caption{Solid lines represent the flux of kinetic energy distribution of antihelium-3 in $pp$ collisions of $\rm \sqrt{s}$ = 7.7 GeV, 17.3 GeV, 53 GeV, and a sum of the three calculations. Pink band and orange band are results from Ref.~\cite{Blum:2017qnn}, and Ref.~\cite{Poulin:2018wzu}, respectively. The red dashed line shows the result derived by ALICE, which is mainly motivated by their new antihelium-3 inelastic cross section data at TeV energy~\cite{ALICE:2022zuz}. The black dash-dotted line indicates the sensitivity of AMS-02 with 5 years of integration time~\cite{Kounine:2010js}.}
    \label{fig:he3flux}
\end{figure}

Figure~\ref{fig:he3flux} presents the flux of antihelium-3 in cosmic rays. Since we can only confirm that the production cross section and momentum distribution of antihelium-3 from our calculations reproduce the experimental data in $pp$ collisions at TeV energies~\cite{ALICE:2017xrp}. Per model calculations, the cross section of antihelium-3 increases with collision energies and reaches its maximum value at high energies~\cite{Zhao:2020irc}. Our calculation can be regarded as an upper limit of our study. We note that our upper limit is lower than the calculation in Ref.~\cite{Blum:2017qnn} (pink band), and is higher than the result in Ref.~\cite{Poulin:2018wzu} (orange band). Since we follow the procedure for treating cosmic ray propagation in Ref.~\cite{Blum:2017qnn}, the flux difference between ours and the Ref.~\cite{Blum:2017qnn} are mainly from the production cross section or the detailed coalescence treatment where our calculation relies on the dynamical distributions of antiprotons and antineutrons. Also, different antihelium fluxes have been obtained with hadronization models EPOS-LHC and DPMJET~\cite{Ding:2018wyi}. For quantitative estimation on the uncertainty level, we calibrate the calculation at low kinetic energy with the antihelium-3 cross section data for a 70 GeV proton hitting on an Al target, where the antihelium-3 to antideuteron ratio is presented in Ref.~\cite{Antipov:1971iq}. Antihelium-3 flux obtained in this way is found to be lower than the upper limit at low kinetic energy by one order of magnitude.

In Ref.~\cite{ALICE:2022zuz}, the ALICE collaboration has reported their calculation of antihelium-3 flux, which is corrected for the absorption cross section of antihelium-3 in the ALICE detector. This estimate, shown as the red dashed line, is lower than the one of Ref.~\cite{Blum:2017qnn} and is close to our result at high kinetic energy (T $\geqslant$ 10 GeV/n). However, our result is about one to two order of magnitude higher than the ALICE results at low kinetic energy (T $\leqslant$ 10 GeV/n). The absorption of antihelium-3 occurs mainly in the low kinetic energy region~\cite{ALICE:2022zuz} and is less than a factor of 2 in the currently discussed kinematic range. It is unlikely that the difference between our result and that from Ref~\cite{ALICE:2022zuz} is due to absorption. It suggests that the antihelium-3 from background $pp$ collisions in cosmic rays is large in low kinetic energy region. Nevertheless, our result is lower than the projected sensitivity of the AMS-02 after 5 years of operation~\cite{Kounine:2010js} by one to two order of magnitude. By comparison with the antihelium-3 fluxes from dark matter predicted in model calculations~\cite{vonDoetinchem:2020vbj}, our result is lower than theirs by one to three order of magnitude in low kinetic energy region. These arguments support the idea to search for dark matter signal by antinucleus detection in the space in the low kinetic energy region. Further systematic study on the existing AMS-02 events~\cite{AMSantihe} or future experiments with detector upgrades such as AMS-100~\cite{Schael:2019lvx} or GAPS~\cite{Aramaki:2014oda} are promising. Measurements of antinucleus production in low energy $pp$ collisions and understanding the detailed production mechanism are important ingredients for improving the predictive power of secondary antnucleus flux in cosmic rays.

\section{Summary}
In summary, the flux distributions of antiproton, antideuteron, and antihelium-3 from the background $pp$ collisions in cosmic rays are studied. The AMPT model is applied to generate the momentum spectra and spatial distributions of antiproton and antineutron in $pp$ collisions. It is tuned to reproduce the $p_{\rm T}$ spectra of proton and antiproton in $pp$ collisions at CERN Large Hadron Collider energies, and the $p_{\rm T}$ and rapidity distributions of antiproton at CERN Super Proton Synchrotron energies. A dynamical coalescence model is then applied to calculate the spectra of antideuteron and antihelium-3. It is found that these calculations describe the $p_{\rm T}$ spectra of antideuteron at LHC energies and the yield ratio of antideuteron to antiproton over a broad energies. Finally, the CR grammage model with solar modulation is used to estimate the antinucleus fluxes in cosmic rays. With successful reproduction of the antiproton flux measured by space detection, our prediction for antideuteron flux is consistent with calculations in the literature. However, our upper limit of antihelium-3 flux sits in-between different estimations, and is lower than the projected sensitivity of AMS-02 experiment by 1 to 2 order of magnitude. Our study of potential background sources from hadronic processes supports the search for dark matter signature via detection of antinuclei in cosmos.

\section{acknowledgements}
Discussion with Dr. Rui Wang is appreciated. This work is support in part by the Strategic Priority Research Program of Chinese Academy of Sciences, Grant No. XDB34030000, by the Guangdong Major Project of Basic and Applied Basic Research No. 2020B0301030008, by the National Natural Science Foundation of China under Contract No. 12025501, No. 11890710, No. 11890714, No. 12147101, the U.S. DOE Office of Science under contract Nos. DE-SC0012704, DE-FG02-10ER41666, and DE-AC02-98CH10886.


\bibliography{myref}

\end{document}